\documentstyle[12pt]{article}
\begin{document}
\vspace{2.5cm}

\hspace{9cm} JHU-TIPAC-930003

\hspace{9cm} January 1993

\vspace{2cm}

\begin{center}{\bf FERMION DENSITY INDUCED INSTABILITY OF THE W-BOSON PAIR
CONDENSATE IN STRONG MAGNETIC FIELD}
\footnote{This work has been
supported by the U.S.
National Science Foundation, grant PHY-90-9619.}
\end{center}

\vspace{1cm}

\begin{center}
Erich R. Poppitz\footnote{E-mail: POPPITZ@CASA.PHA.JHU.EDU}
\end{center}

\vspace{.5cm}

\begin{center}

{\it Department of Physics and Astronomy}

{\it The Johns Hopkins University}

{\it Baltimore, MD 21218, USA}

\end{center}

\vspace{1cm}

\centerline {ABSTRACT}
\vspace{.5cm}

The electroweak vacuum structure
in an external magnetic field close to the
lower critical value is considered at finite fermion density.
It is shown that the leading effect of the fermions is
to reduce the symmetry of the $W$-pair condensate in the direction
of the magnetic field.
The energy is minimized by the appearance of a helicoidal
structure of the condensate along the magnetic field.

\newpage

{\bf 1.} In the presence of strong external magnetic fields
the ground state of the Weinberg-Salam theory exhibits a very
interesting structure. When the field is increased above
the critical value $H_{c1}=m_W^2/e\sim 10^{24}$ Gauss,
$W^+$-$W^-$ pairs condense in an Abrikosov-type
vortex lattice. Increasing the field above $H_{c2}=H_{c1}/{\rm cos}\theta_W$
leads to the vanishing of the Higgs vacuum expectation value
and a restoration of ${\rm SU}(2)\times{\rm U}(1)$-symmetry.
It is thus possible to
reach the ordered phase of the electroweak theory not only by increasing the
temperature
but also by increasing the magnetic field \cite{ambjorn1},
\cite{ambjorn2}, \cite{ambjorn5}.

Previously, the
development of the instability for $H>H_{c1}$ has been studied
by neglecting the presence of fermions in the system. For gauge and
Yukawa couplings in the perturbative regime,
the effect of the Dirac sea
will be to renormalize the effective action of the bosonic fields. The
presence of a finite density of fermions, however, gives a qualitatively new
effect on the intermediate phase, because of the fermion number nonconservation
in the Weinberg-Salam theory.

The lattice of vortices, found in \cite{ambjorn1},
is homogeneous in the direction of the external field.
In the presence of finite
fermion density, however, the development of an inhomogeneity in the field
direction
causes (through the $B+L$-anomaly) some occupied positive energy levels
to fall into the Dirac sea, lowering the Fermi energy. This energy
gain may overcome the energy loss due to the inhomogeneity,
leading to the formation of a structure that is periodic in the
direction of the field. The physics of this instability is reminiscent
of the Peierls transition in 1-d metals \cite{peierls}, or more closely,
of the instability of the charged-$W$ condensate in cold nonneutral
fermionic matter \cite{rubakov1}.

This letter is organized as follows. In sect.2 we review the essential
properties of the Ambj\o rn-Olesen solution. We show that the fermion
number anomaly leads, at finite chemical potential, to an instability
towards the formation of a helicoidal structure of the $W$-pair
condensate along the field direction. In sect.3 we discuss the fermionic
ground state at $H\simeq H_{c1}$ and for fermion masses and chemical
potential obeying
$m_f\ll \mu\ll\sqrt{eH}$. For fields close to the lower critical value,
the state in a homogeneous magnetic field
can then be treated as the unperturbed
ground state, and the $W$-pair condensate as a perturbation.
We show that under these conditions
the leading effect of the fermions is the destabilization
of the $W$-condensate towards inhomogeneity in the field direction. Sect.4
contains discussion of the results and possible applications.

\vspace{1cm}

{\bf 2.} Let us consider first the bosonic sector of the standard model.
The static energy functional in the unitary gauge is \cite{macdowell}:
$$
{\cal E}_0 = \frac{1}{4}\left(\partial_{[i}W^3_{j]}\right)^2
           +\frac{1}{4}\left(\partial_{[i}B_{j]}\right)^2
           +\frac{1}{2}\vert \partial_{[i}W^{\dagger}_{j]}\vert^2
$$
\begin{equation}
\label{b}
        -{ i g \over 2} \partial_{[i} W^3_{j]} W^{\dagger}_i W_j
          -\frac{g^2}{2}\left[\vert W^{\dagger}_i W^{\dagger}_i\vert^2
                          -\left(W^{\dagger}_i W_i\right)^2\right]
\end{equation}

$$
+ i g\partial_{[i}W^{\dagger}_{j]} W^3_i W_j
      - i g \partial_{[i}W_{j]}W^3_i W^{\dagger}_j
      -g^2 W^3_i W^{\dagger}_jW^3_{[i}W_{j]}
$$
$$
+\left(\vec{\nabla}\phi\right)^2 + \lambda\left(\phi^2 - v^2\right)^2
+{g^2 \over 2} \phi^2 W_i^{\dagger}W_i
+{g^2 \over 4{\rm cos}\theta} \phi^2 Z_i Z_i .
$$
Here
$W_i\equiv(W_i^1-iW_i^2)/\sqrt{2}$ is the
charged-$W$ field,
$Z_i=W^3_i {\rm cos}\theta - B_i {\rm sin}\theta $,
the electromagnetic vector potential is
$A_i= W^3_i {\rm sin}\theta +  B_i {\rm cos}\theta $, and
$\phi$ is the Higgs scalar.

In
the presence of a static homogeneous magnetic field  $H>H_{c1}$,
pointing in the third direction,
the ground state is characterized by nonvanishing
$W$, $Z$, $A$, $(\phi -v)$-condensates.
For the static solution, found by Ambj\o rn
and Olesen \cite{ambjorn1}
\begin{equation}
\label{a0}
A_3 =W_3 =Z_3 =0,\hspace{.3cm} A_0 =W_0 = Z_0 =0 ,
\end{equation}
and
\begin{equation}
\label{a3} W_2 =iW_1 \equiv W(x_1, x_2).
\end{equation}
The condensates are homogeneous along the third axis and periodic
in the plane perpendicular to the magnetic field.
The vortex-lattice structure has a unit cell with area
$S_k = 2\pi k /m_W^2$, where $k$
is the winding number of the phase of the $W$-field
along the cell boundary. The
$W$-field has $k$ zeros inside the cell. The
energy is minimized for $k=1$ and a hexagonal lattice  \cite{macdowell}.

The properties (\ref{a0}),(\ref{a3}) of the
$W$-condensate are sufficient to demonstrate that
a finite density of fermions causes it to be unstable
towards inhomogeneity in the field direction.
In the presence of finite fermion density the fermion number
anomaly in the standard model
leads to the appearance of a Chern-Simons term in the bosonic
effective action \cite{redlich}
\begin{equation}
\label{ecso}
{\cal E}_{CS}=\mu f n_{CS},
\end{equation}
where $f$ is the number
of doublets, and
\begin{equation}
\label{ncs}
n_{CS}={g^2 \over 8\pi^2}\left[ \epsilon^{ijk}W^a_i\partial_j W^a_k
+{g \over 3}\epsilon^{ijk}\epsilon^{abc} W^a_i W^b_j W^c_k
- {\rm tan}^2\theta \epsilon^{ijk}B_i\partial_j B_k\right].
\end{equation}
To understand the appearance of (\ref{ecso}) in the effective action,
it is useful to note that the introduction of the chemical potential
is equivalent to coupling an external "fermion number gauge field"
${\cal F}_{\alpha}$ to the fermions, such that
${\cal F}_{\alpha}=(\mu, 0, 0, 0)$. Then calculating the term in the
effective action, linear in $\mu$ and quadratic in the background,
is equivalent to computing the correlator
$$
\langle J_{\alpha}^f J_{\beta}^a J_{\gamma}^b \rangle ,
$$
where $J^f$ is the fermion number current, $J^a$, $J^b$ are
SU(2) and U(1) currents. The totally antisymmetric contribution to this
correlator is determined by the anomaly equation:
$$
\partial_{\mu} J^f_{\mu}= {f g^2 \over 32 \pi^2}
\left( F_{\mu \nu}^a {\tilde F}^a_{\mu \nu} - {\rm tan}^2 \theta
       F_{\mu\nu} {\tilde F}_{\mu\nu}\right),
$$
($F^a$, ${\tilde F}^a$ and $F$, ${\tilde  F}$ are the SU(2) and U(1) field
strengths and their duals, respectively) and coincides
with (\ref{ecso}).

{}From (\ref{a0}),(\ref{a3})
it follows that (\ref{ncs}) reduces to:
\begin{equation}
\label{cs}
{\cal E}_{CS}=\mu f {g^2 \over 8\pi^2}\left[
\epsilon^{ij3}W^{\dagger}_i\partial_3 W_j + h.c.
+  \epsilon^{ij3}W^3_i\partial_3 W^3_j
- {\rm tan}^2 \theta \epsilon^{ij3}B_i\partial_3 B_j\right] .
\end{equation}
We see that the appearance of an inhomogeneity in the third direction
leads to nonzero density of the Chern-Simons number of the condensate.
$n_{CS}$ is equal to the number density of the positive energy fermion levels
which have fallen into the Dirac sea. Note that for the original solution
(\ref{a0}),(\ref{a3}) $n_{CS}=0$.
To find the form of the condensate
which minimizes the energy we have to consider the total bosonic energy
functional, which includes the energy loss due to the inhomogeneity:
\begin{equation}
\label{etotal}
{\cal E}_{total}={\cal E}_0 + {\cal E}_{CS},
\end{equation}
where ${\cal E}_0$ is given by (\ref{b}).
We see that an $x_3$-dependent isospin rotation
of the $W$-condensate around the third axis
\begin{equation}
\label{inhom}
W_1\rightarrow e^{ik_3 x_3}W_1\equiv -ie^{ik_3 x_3}W,
\hspace{.3cm} W_2\rightarrow e^{ik_3 x_3}W_2 \equiv e^{ik_3 x_3}W,
\end{equation}
does not affect ${\cal E}_0$ except for the term with
two derivatives (recall that for the Ambj\o rn-Olesen solution
all components along the third axis vanish). This two-derivative
contribution represents the "elasticity" energy loss due to the
inhomogeneity.
On the other hand, the ${\cal E}_{CS}$ contribution
is linear in $k_3$, the total energy change being:
\begin{equation}
\label{balance}
\delta{\cal E}=k_3^2W^{\dagger}W -
\mu f \frac{g^2}{8\pi^2}k_3 W^{\dagger}W .
\end{equation}
Clearly, for small enough $k_3$ the linear term dominates
and $\delta {\cal E}$ is negative and minimized for a nonzero value of
$k_3$:
\begin{equation}
\label{k3}
k_3={g^2 \mu f \over 16\pi^2} .
\end{equation}
The last two expressions show that the finite density of fermions
causes an instability of the vortex lattice towards inhomogeneity
in the direction of the magnetic field and
leads to the formation of a helicoidal structure of the $W$-condensate.

Being determined by the exact anomaly equation the
Chern-Simons contribution ${\cal E}_{CS}$ is valid for all momenta
(well below the UV cut-off) of the background fields. Therefore the
density-induced instability has a quite general character.

For arbitrary fermion densities and fields much stronger
than the lower critical one, the $W$-condensate is not small
and we are not aware of the nature of the fermionic ground state.
In particular,
then we cannot insist that  the
most important influence of the fermions on the condensate
is the appearance of the inhomogeneity.

\vspace{1cm}

{\bf 3.} Let us therefore consider magnetic field strengths
close to the lower critical value $H_{c1}$ and show that
the instability described in the previous section
is then the leading effect of the fermions.

Let us decompose $A=\bar{A} + A^{\prime}$, where $\bar{A}$
is the electromagnetic vector potential,
corresponding to the homogeneous part of the
magnetic field background,
and introduce the small parameter
 $\epsilon = (H-H_{c1})/H_{c1} \ll 1$.
For such fields the $A^{\prime}, W, Z, \phi - v$ condensates
are small \cite{macdowell}:
$$
\vert W \vert \sim \epsilon v
$$
$$
A^{\prime}, Z, \phi -v \sim \epsilon^2 v.
$$
Their characteristic momenta in
the direction perpendicular to the magnetic field are
\cite{macdowell}:
\begin{equation}
\label{mom}
k^W_{1,2}\sim \epsilon gv ,
\end{equation}
$$
k^W\gg k^{A^{\prime}}, k^Z, k^{\phi}.
$$

Now let us couple fermions to the above background. The radius
of the lowest Landau orbit  \cite{llqm}
of a fermion with charge $eq$, $e>0$, in a constant homogeneous
magnetic field is
$$r_H=\sqrt{2/e\vert q\vert H} .$$
Note that for $H\simeq m_W^2/e$ this is of the order of the
unit cell size $\sqrt{S_1}$, and that $k^W_{1,2} r_H \ll 1$,
which means that the condensates
vary slowly  along the Landau orbit. Therefore we
will consider the state in a homogeneous magnetic field
as the unperturbed
ground state, i.e. treat the constant part of $H$ exactly and
consider $A^{\prime}, W,Z, \phi - v$ as small perturbations.

For chemical potential $\mu$, obeying
\begin{equation}
\label{mu}
m_f\ll \mu\ll\sqrt{eH} ,
\end{equation}
all fermions will occupy the lowest Landau level.
Their number density is then $n_q= e\vert q\vert H\mu / (2\pi^2)$,
and the energy density: ${\cal E}_q=  e \vert q\vert H \mu^2 / (4\pi^2)$.
{}From (\ref{mu}) we see that
the Fermi energy density can be neglected in comparison with the energy
density of the constant magnetic background:
${\cal E}_q \ll {\cal E}_H= H^2 / 2$.

For the fields under consideration the shift of the Dirac sea
energy can also be neglected  \cite{llqed}:
For a fermion
with mass $m_f$ and charge $qe$ in magnetic
field $H \gg  m_f^2 /(\vert q \vert e)$
it equals:
$$
{\cal E}_H^{vac} - {\cal E}_{H=0}^{vac} =
{\vert q\vert^2 e^2 H^2 \over 24 \pi^2}{\rm ln}
{\vert q\vert e H \over m_f^2}
$$
and becomes comparable to ${\cal E}_H = H^2 /2 $ only for field
strengths of the order of
$${m_f^2 \over \vert q\vert e}
e^{12\pi^2 \over \vert q \vert^2 e^2} \gg H_{c1}={m_W^2 \over e} .$$

Finally, higher derivative terms in the effective bosonic action
can be neglected if the momenta of the background in the
direction perpendicular to $H$ obey $k_{1,2}\ll \sqrt{eH}$,
and the ones along $H$: $k_3 \ll \mu$ \cite{rubakov1}.
The first condition is obeyed
for fields close to the lower critical one,
as follows from (\ref{mom}); the second is seen to hold from (\ref{k3}).

The above considerations show that
for small coupling and field strengths $\sim H_{c1}$
quantum
corrections due to fermions lead to renormalization of the coefficients
of the bosonic energy functional only.
Therefore, the Chern-Simons term generated by the anomaly
is the leading contribution to the bosonic effective action,
and the ground state is characterized by the helicoidal structure
(\ref{inhom}) with wave vector (\ref{k3}).

Let us note that we cannot prove that this is the only instability
occuring. All we have shown is that
(\ref{k3}) minimizes the energy within the Ansatz
(\ref{inhom}).

The $x_3$-dependent phase of the $W$-condensate leads to
a current flow along  the third axis, which in its turn will affect
the system, but in the small coupling regime these effects
are suppressed by additional powers of $e$.

The scale of the inhomogeneity $k_3^{-1}$ is much larger than the
unit cell size, for the densities for which the above analysis is
valid. If however (\ref{k3}) is taken at face value, they
become comparable only for $\mu \sim m_W / g^2$.
At these densities however, the system is classically
unstable towards transition to
the vacuum with different fermion number.
Note that in our case
there is no true fermion number violation - if the magnetic
field is turned off the levels appear again and the initial
number of fermions is restored.

\vspace{1cm}

{\bf 4.} Turning to possible applications, we note that the
conditions necessary for the realization of the fermion-induced
instability might be present only in the early Universe.
The values of the critical fields
$H_{c1}$,$H_{c2}$ go to
zero as $T\rightarrow T_c$, where $T_c \sim 100{\rm GeV}$ is the critical
temperature of the electroweak phase transition. Therefore, slightly
below $T_c$, fields much smaller than $10^{24}$ Gauss could drive
the system in the intermediate phase \cite{ambjorn2}.
Such fields might be present
before the phase transition, being generated by quantum fluctuations during
inflation \cite{ratra}.
They could also be created during the electroweak transition itself,
realizing thus a "vacuum dynamo" effect
 \cite{ambjorn2}, \cite{v}.

The result for the fermion generated ${\cal E}_{CS}$ holds at
finite temperature as well, to leading order in the high temperature expansion
$\mu, m_f \ll T$  \cite{redlich}. If there were some "seed" magnetic
field \cite{ratra}
prior to the electroweak transition, then below $T_c$ the mixed phase
might be energetically favourable for some period of time.
As we saw above it has two distinct scales of inhomogeneity -
$(m_W)^{-1}$ in the direction
perpendicular to the magnetic field, and
$4eH /(g^2 n_f) \sim 4 m_W^2 /(g^2 n_f) \gg (m_W)^{-1}$
in the longitudinal direction.
This strong anisotropy would affect the gravitational wave background
\cite{sazhin},
but it is not clear whether
its amplitude is strong enough to lead to
any observable effect.

\vspace{.5cm}

To conclude, we have shown that a finite fermion density
destabilizes the $W$-pair condensate in strong magnetic field
and breaks the translational invariance in the direction
of the field, leading to the appearance of a helicoidal structure.

\vspace{.5cm}

It is a pleasure to acknowledge helpful discussions with
J. Bagger,

G. Feldman and T. Gould.


\newpage

\begin{thebibliography}{99}

\bibitem{ambjorn1} J. Ambj\o rn and P. Olesen, {\it Phys. Lett.} B{\bf 218}
(1989) 67.

\bibitem{ambjorn2} J. Ambj\o rn and P. Olesen, {\it Nucl. Phys.} B{\bf 330}
(1990) 193.

\bibitem{ambjorn5} J. Ambj\o rn and P. Olesen, {\it Nucl. Phys.} B{\bf 315}
(1989) 606.

\bibitem{peierls} R. Peierls, {\it More Surprises in Theoretical Physics},
(Princeton UP, Princeton, NJ, 1991).

\bibitem{rubakov1} D.I. Deryagin, D.Yu. Grigoriev and V.A. Rubakov,
{\it Phys. Lett.} B{\bf 178} (1986) 385.

\bibitem{macdowell} S.W. MacDowell and O. T\"ornkvist, {\it Phys. Rev.}
D{\bf 45} (1992) 3833.

\bibitem{redlich} A.N. Redlich and L.C.R. Wijewardhana, {\it Phys. Rev. Lett.}
{\bf 54} (1985) 970;

K. Tsokos, {\it Phys. Lett.} B{\bf 157} (1985) 413.

\bibitem{llqm} L.D. Landau and E.M. Lifshitz, {\it Quantum Mechanics},
3rd ed., (Pergamon Press, NY, 1977).

\bibitem{llqed} V.B. Berestezkii, E.M. Lifshitz and L.P. Pitaevskii,
{\it Quantum Electrodynamics}, 2nd ed., (Pergamon Press, NY, 1982).

\bibitem{ratra} B. Ratra, Caltech preprints CALT-68-1750 and CALT-68-1751,
October 1991.

\bibitem{v} T. Vachaspati, {\it Phys. Lett.} B{\bf 265} (1991) 258;

P. Olesen, {\it Phys. Lett.} B{\bf 281} (1992) 300.

\bibitem{sazhin} D.I. Deryagin, D.Yu. Grigoriev, V.A. Rubakov and
M.V. Sazhin, {\it Mod. Phys. Lett.} A{\bf 1} (1986) 593.
\end{thebibliography}
\end{document}